\documentclass[aps,prd,onecolumn,superscriptaddress,floats,amsfonts,amsmath,amssymb,linenumbers]{revtex4}
\usepackage{graphicx}
\usepackage{epstopdf}
\usepackage[active]{srcltx}
\usepackage{bm}
\usepackage[]{latexsym}

\usepackage{color}

\usepackage{float}

\usepackage{hyperref}
\hypersetup{pdfauthor={huchiayu},pdftitle={},
            pdfsubject={},pdfcreator={pdfLaTeX with hyperref (\today)}}

\usepackage{lineno}
\begin{document}

\restylefloat{figure}

\title{Near-Field Effects of Cherenkov Radiation Induced by Ultra High Energy Cosmic Neutrinos}


\author{Chia-Yu Hu}
\email{r96244004@ntu.edu.tw}
\affiliation{Graduate Institute of Astrophysics, National Taiwan University, Taipei, Taiwan 10617}
\affiliation{Leung Center for Cosmology and Particle Astrophysics (LeCosPA), National Taiwan University, Taipei, Taiwan, 10617}

\author{Chih-Ching Chen}
\email{chen.chihching@gmail.com}
\affiliation{Graduate Institute of Astrophysics, National Taiwan University, Taipei, Taiwan 10617}
\affiliation{Leung Center for Cosmology and Particle Astrophysics (LeCosPA), National Taiwan University, Taipei, Taiwan, 10617}

\author{Pisin Chen}
\email{chen@slac.stanford.edu}
\email{pisinchen@phys.ntu.edu.tw}
\affiliation{Graduate Institute of Astrophysics, National Taiwan University, Taipei, Taiwan 10617}
\affiliation{Leung Center for Cosmology and Particle Astrophysics (LeCosPA), National Taiwan University, Taipei, Taiwan, 10617}
\affiliation{Department of Physics, National Taiwan University, Taipei, Taiwan 10617}
\affiliation{Kavli Institute for Particle Astrophysics and Cosmology, SLAC National Accelerator Laboratory, Menlo Park, CA 94025, U.S.A.}


\begin{abstract}
The radio approach for detecting the ultra-high energy cosmic neutrinos has become a mature field.
The Cherenkov pulse in radio detection originates from the charge excess of particle showers due to Askaryan effect.
The conventional way of calculating the Cherenkov pulse by making far-field approximation fails when the size of elongated showers become comparable with detection distance.
We investigate the Cherenkov pulse in near-field by a numerical code based on the finite-difference time-domain (FDTD) method.
Our study shows that the near-field radiation exhibits very different behaviors from the far-field one and therefore can be easily recognized.
For ground array neutrino detectors,
the near-field radiation would provide a unique signature for ultra high energy electromagnetic showers induced by the electron neutrino charge-current interaction.
This can be useful in neutrino flavor identification.

\vspace{3mm}

\end{abstract}
\maketitle

\section{Introduction}
Cosmic neutrinos, as a probe of the universe to the highest energy regime, are indeed wonderful in many aspects.
Due to their extremely small interaction cross section, 
they can penetrate through galactic infrared (IR) and cosmic microwave background (CMB) photons, 
while photons of energy above 10 TeV would be attenuated.
Furthermore, being uncharged, they propagate along straight lines and are therefore able to point directly back to their sources,
while protons or other charged particles would be deflected by the magnetic fields in the universe.

Ultra-high energy cosmic rays (UHECRs) have been observed up to $\approx$ $10^{19.6}$ eV.
The source of such amazingly energetic events remains a mystery.
Above this energy scale, UHECRs interact with CMB photons through the Greisen-Zatsepin-Kuzmin(GZK) processes~\cite{gzk_process}.
The GZK cut-off of the cosmic ray energy spectrum was first observed 
by the High Resolution Fly's Eye Experiment~\cite{cutoffHiRes}
and later confirmed by the Pierre Auger Observatory~\cite{cutoff}. 
As such, the corresponding GZK neutrinos,
a necessary by-product of the GZK process,
are almost guaranteed to exist.
Nevertheless, none of these have been observed so far. 
Detection of the GZK neutrinos would provide critical information 
for unraveling the mystery of the origin and evolution of cosmic accelerators, 
and will be one of the most exciting prospects in the coming decade~\cite{pisin_whitepaper}.

A promising way of detecting UHE neutrinos is the radio approach.
When an ultra-high energy cosmic neutrino interacts with ordinary matters on the Earth, 
it would lead to a hadronic debris, caused by either charged-current or neutral-current weak interaction.
The former also produces a lepton with corresponding flavor.
Both the high energy leptons and hadronic debris induce particle showers.
As proposed by Askaryan in the 1960's~\cite{Askaryan}, 
the high energy particle shower developed in a dense medium would have net negative charges.
This charge imbalance appears as a result of the knocked-off electrons being part of the shower,
as well as positrons in the shower annihilating with electrons of the medium.
The net charges of the showers, typically $20 \%$ of total shower particles, 
serve as a source which emits the Cherenkov radiations when they travel in the medium.
The sizes of the showers are quite localized (tens of cm in radial and a few meters in longitudinal development) 
compared to those developed in the air (km scale),
and therefore result in coherent radiations for wavelengths longer than the shower sizes.
The corresponding coherent wavelength turns out to be in the radio band, from hundreds of MHz to a few GHz.
This Askaryan effect has been confirmed in a series of experiments at the Stanford Linear Accelerator Center (SLAC), 
where different dense media such as silica sand, rock salt and ice were used~\cite{aska1, aska2, aska3, aska4}.

Various  experiments have been proposed based on the idea of Askaryan:
the balloon-borne antenna array (e.g. ANITA~\cite{anita1, anita2}),
the ground-based antenna array buried in salt (e.g. SalSA~\cite{salsa}) or ice (e.g. RICE~\cite{rice}),
and the radio telescope searching for lunar signals (e.g. LUNASKA~\cite{lunaska}).
As ground-based experiments have the advantage of low noise and low energy threshold,
they play an important role in the next-generation experiments 
aimed at detecting some tens of GZK neutrinos per year.
However, for the extremely high energy neutrino event, especially for the electron neutrino, 
it is very likely that the longitudinal size of the shower would become comparable to the distance between the antenna and the shower.
Under this situation, the common assumption of far-field radiation does not apply and near-field effects become nonnegligible.
In this paper, we study the impact of the near-field effect to the radiation pattern via a numerical method.
The organization of this paper is as follows:
Section II will discuss the underlying physics of the shower elongation.
Section III will introduce the numerical method and our setup.
Section IV will present the implementation of parallel computing to gain a satisfying efficiency of our numerical calculation.
In Section V, we analyse the results in both time-domain and frequency-domain.
We discuss the features of the near-field radiation pattern 
and compare the far-field pattern between our results and the theoretical ones for validation.

\section{Coherent Cherenkov Pulses}
In the radio detection experiment, 
signals come from Cherenkov radiations of net charges in the shower.
The key concept which makes this technique possible is coherent emission.
In fact, Cherenkov radiation is a broad band emission and the intensity increases as frequency increases.
For a single charged particle,
the Cherenkov signal in the radio band should be the weakest in the spectrum.
The compact size of the shower is what makes the radio signal so special. 
Coherent emission greatly enhances the signal strength in the radio band.

The electric field and its associated vector potential of the Cherenkov radiation can be calculated by solving inhomogeneous Maxwell equations, 
as demonstrated in the paper of Zas, Halzen and Stanev~\cite{ZHS}.
The vector potential can be obtained by the Green's function method:
\small{
\begin{equation}
	\vec{A}_{\omega}(\vec{x}) 
	= \frac{1}{4 \pi \epsilon_0 {\rm c}^2}
	\int\!\!d^{3}\!x'\,\frac{\exp{(ik|\vec{x}-\vec{x}\,'|)}}
	{|\vec{x}-\vec{x}\,'|}\int\!\!dt'\,\exp{(i\omega t')}
	\vec{J}(t',\vec{x}\,'), 
	\label{eq:GreenMethod}
\end{equation}
}
\normalsize
where $k$ is the wavenumber,
$\vec{x}$ is the position of the detector,
$\vec{x}\,'$ is the position of the shower particles 
and $\vec{J}$ is the current source.
Adopting the Fraunhoffer approximation
\small{
\begin{equation}
	\frac{ \exp{(ik|\vec{x}-\vec{x}\,'|)} }{|\vec{x}-\vec{x}\,'|}
	\approx   \frac{ \exp{(ik|\vec{x}|-i k\hat{x}\cdot\vec{x}\,')} }{R} ~,
	\label{eq:fraun_appr}
\end{equation}
}
\normalsize
where $R$ is the absolute value of $\vec{x}$,
the integration in Eq.(~\ref{eq:GreenMethod}) can be greatly simplified and 
therefore enhances the computational efficiency in the Monte Carlo simulation~\cite{ZHS, AZ_1d, AZ_unified, AZ_thinned}.
The validity of this approximation relies on several length scales:
the detection distance ($R$), the spatial size of the shower ($l$) and the wavelengths of interest ($\lambda$).
The Fraunhoffer approximation works well under the condition
\small{
\begin{equation}
	\frac{l^2}{\lambda R}~\rm{sin}^2\theta \ll 1,
	\label{eq:farfield_condition}
\end{equation}
}
\normalsize
where $\theta$ is the angle between the shower axis and the observational direction.

However, for ultra-high energy showers, the longitudinal development is longer,
especially for electromagnetic showers that suffer from Landau-Pomeranchuk-Migdal (LPM) suppression~\cite{LPM, lpm_klein}.
Electromagnetic showers can be produced by charge current generated leptons.
For an electromagnetic shower of EeV-scale energy,
the impact of LPM effect on shower development has been investigated by Monte Carlo simulations~\cite{Niess, Bolmont, AZ_thinned}.
Electromagnetic showers of primary energy $10^{20}$ eV can be extended to about 200-m long with great fluctuations.
In such cases, the far-field condition cannot be satisfied for distances up to several kilometers,
Meanwhile, the typical detection distance for ground array detectors is about 1 km, as dictated by the attenuation length of radio signals in ice.
In addition,
the separation between detection stations in the proposed layout is also about 1 km,
which makes the possible detection distance even shorter.
Under these circumstances, the Fraunhoffer approximation is invalid 
and one has to deal with the complicated integration in Eq.(~\ref{eq:GreenMethod}).

\begin{figure}[H]
	\begin{center}
	\includegraphics[width=9cm, height=6cm ]{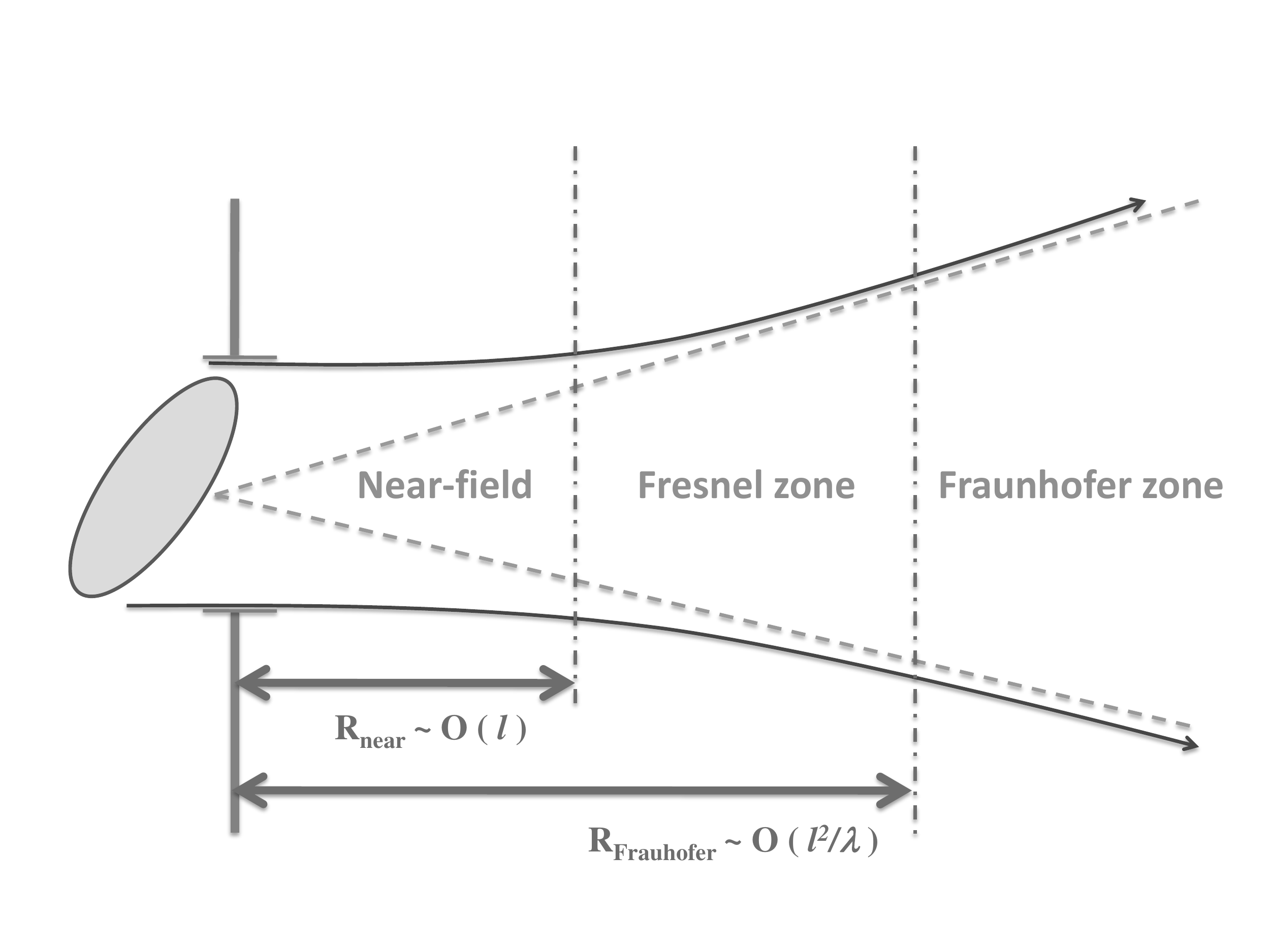}
	\caption{A cartoon picture that illustrates the diffraction phenomenon in three different radiation zones. In near-field there is little diffraction, but the radiation sweeping zone is as wide as the longitudinal development of the shower. In Fresnel zone the diffraction starts to take effect and part of the radiation leaks out of the Cherenkov angle.}
	\label{fig:near-field sweeping zone}
	\end{center}
\end{figure}

In the paper of Buniy and Ralston~\cite{BR},
the correction has been made by the saddle-point approximation,
which deals with Fresnel zone radiation where the Fraunhoffer condition fails.
However, it still can not cope with extreme cases for $R \sim l$, i.e. the near-field radiation, 
which is an inevitable concern for showers of hundred meters long.
We address this problem by a numerical method based on first principle
so that near-field radiations can be efficiently obtained.
Figure (\ref{fig:near-field sweeping zone}) illustrates the relation among the three different radiation zones.
The Cherenkov radiation can be viewed as an analogy with the diffraction problem in optics.
Although the radiation in near-field is confined within the Cherenkov direction where the diffraction has yet to occur,
it does have a finite width which is at least as wide as the longitudinal development of the shower.
After reaching far-field, 
the radiation starts to diffract out of the Cherenkov angle.
While the far-field approximation is able to correctly deal with the diffraction phenomenon in far-field,
it fails to describe the situation in near-field.

Hadronic showers, on the other hand, are less affected by the LPM effect~\cite{alvarez98, acorne},
since the sources of electromagnetic components in hadronic showers are the decay of neutral pions,
which tend to interact with the medium instead of decay at energy levels above 6.7 PeV.
The far-field condition is well fulfilled for hadronic showers in most practical cases.


\section{Numerical Method}
Numerical algorithms for calculating electromagnetic fields have been in existence for decades. 
However, it was not until the recent rapid growth of computational power that this approach became wildly adopted.
Among all the existing algorithms, 
the finite-difference time-domain (FDTD hereafter) method has several advantages:

\begin{itemize}
\item
It is exceptionally simple to be implemented by computer programs.
\item 
The time-domain approach is well suitable for an impulse signal;
A broad band impulse can be calculated in one single run.
\item 
The algorithm itself is inherently parallel and efficiency can be largely improved via parallel computing.
\end{itemize}

The idea of FDTD was first proposed by Yee in the 1960's~\cite{Yee},
and has been in use for many years for electromagnetic impulse modeling.
Like most numerical finite difference methods, 
space is discretized into small grids and fields are calculated on each grid by solving Maxwell equations.
Adopting a special lattice arrangement (known as the Yee lattice), 
the E-field and H-field are staggered in both space and time and can be calculated in a leapfrog time-marching way.
In the remainder of this section, we review the algorithm of the FDTD method and our setup.

\subsection{Algorithm}
The Maxwell curl equations in differential forms are
\small{
\begin{equation}
	\nabla \times \mathbf{E} = -\mu      \frac{\partial \mathbf{H}}{\partial t} ~,
\end{equation}
\begin{equation}
	\nabla \times \mathbf{H} =  \epsilon \frac{\partial \mathbf{E}}{\partial t} + \sigma \mathbf{E} ~.
\end{equation}
}
\normalsize

The FDTD method approximates derivatives by finite differences. 
The central difference is adopted to achieve 2nd order accuracy in both spatial and temporal derivatives.
We assume cylindrical symmetry along the shower axis (defined as z-axis), 
and therefore all the derivatives with respect to $\phi$ vanish.
In addition, due to the polarization property of Cherenkov radiations, 
the $H_r, H_z, E_\phi$ components also vanish.
This can save large amounts of computer memories and calculation time.
Figure(~\ref{fig:cylind_grid}) shows the configuration of the lattice under these assumptions.
Maxwell equations in a cylindrical coordinate then reduce to:
\small{
\begin{eqnarray}
	\frac{\partial E_r}{\partial z} - \frac{\partial E_z}{\partial r} = -\mu \frac{\partial H_\phi}{\partial t}, \\
	-\frac{\partial H_\phi}{\partial z} = \epsilon \frac{\partial E_r}{\partial t}, \\
	\frac{1}{r} \frac{\partial (rH_\phi)}{\partial r} = \epsilon \frac{\partial E_z}{\partial t}.  
\end{eqnarray}
}
\normalsize
where we have set $\sigma = 0$, assuming lossless medium.

The above equations can be discretized in both space and time 
with a spatial interval of $\Delta r$ and $\Delta z$ in the r and z direction respectively, 
and with a time interval of $\Delta t$.
For example, the discretized field for $E_r(r = i \Delta r, ~z = j \Delta z, ~t = n\Delta t)$ is denoted as $E_r^n(i, j)$.
Replacing derivatives with finite differences, we have
\small{
\begin{eqnarray}
	&&H_\phi^{n+1/2}(i+1/2, j+1/2) = H_\phi^{n-1/2}(i+1/2, j+1/2) \nonumber \\
	&&+ \frac{\Delta t}{\mu \Delta r}[ E_z^n(i+1, j+1/2) - E_z^n(i, j+1/2) ] \nonumber \\
	&&- \frac{\Delta t}{\mu \Delta z}[ E_r^n(i+1/2, j+1) - E_r^n(i+1/2, j-1/2) ] \label{eq:fdtd_Hphi}, \\
	\nonumber \\
	&&E_r^{n+1}(i+1/2, j) = E_r^{n}(i+1/2, j) \nonumber \\
	&&-  \frac{\Delta t}{\epsilon \Delta z}[ H_\phi^{n+1/2}(i+1/2, j+1/2) \nonumber \\
	&&- H_\phi^{n+1/2}(i+1/2, j-1/2) ] \label{eq:fdtd_Er}, \\ 
	\nonumber \\
	&&E_z^{n+1}(i, j+1/2) = E_z^{n}(i, j+1/2) \nonumber \\
	&&+ \frac{\Delta t}{\epsilon r_i \Delta r}[ r_{i+1/2}H_\phi^{n+1/2}(i+1/2, j+1/2) \nonumber \\
	&&- r_{i-1/2}H_\phi^{n+1/2}(i-1/2, j+1/2) ] \label{eq:fdtd_Ez}. 
\end{eqnarray}
\normalsize
Iteratively, we can use Eq.~\ref{eq:fdtd_Hphi} to update the H-field, 
and then use Eq.~\ref{eq:fdtd_Er} and Eq.~\ref{eq:fdtd_Ez} to update the E-field, and so on.
The relation between spatial and temporal grid sizes have been chosen as 
$\Delta r = \Delta z = \delta$, ${\rm c_{ice}}\Delta t = \delta/2$ such that numerical stability is satisfied.
Numerical dispersion can be controlled at an acceptable level if we define the grid size fine enough~\cite{Courant, Taflove}.

\begin{figure}[htbp]
	\begin{center}
	\includegraphics[width=8cm]{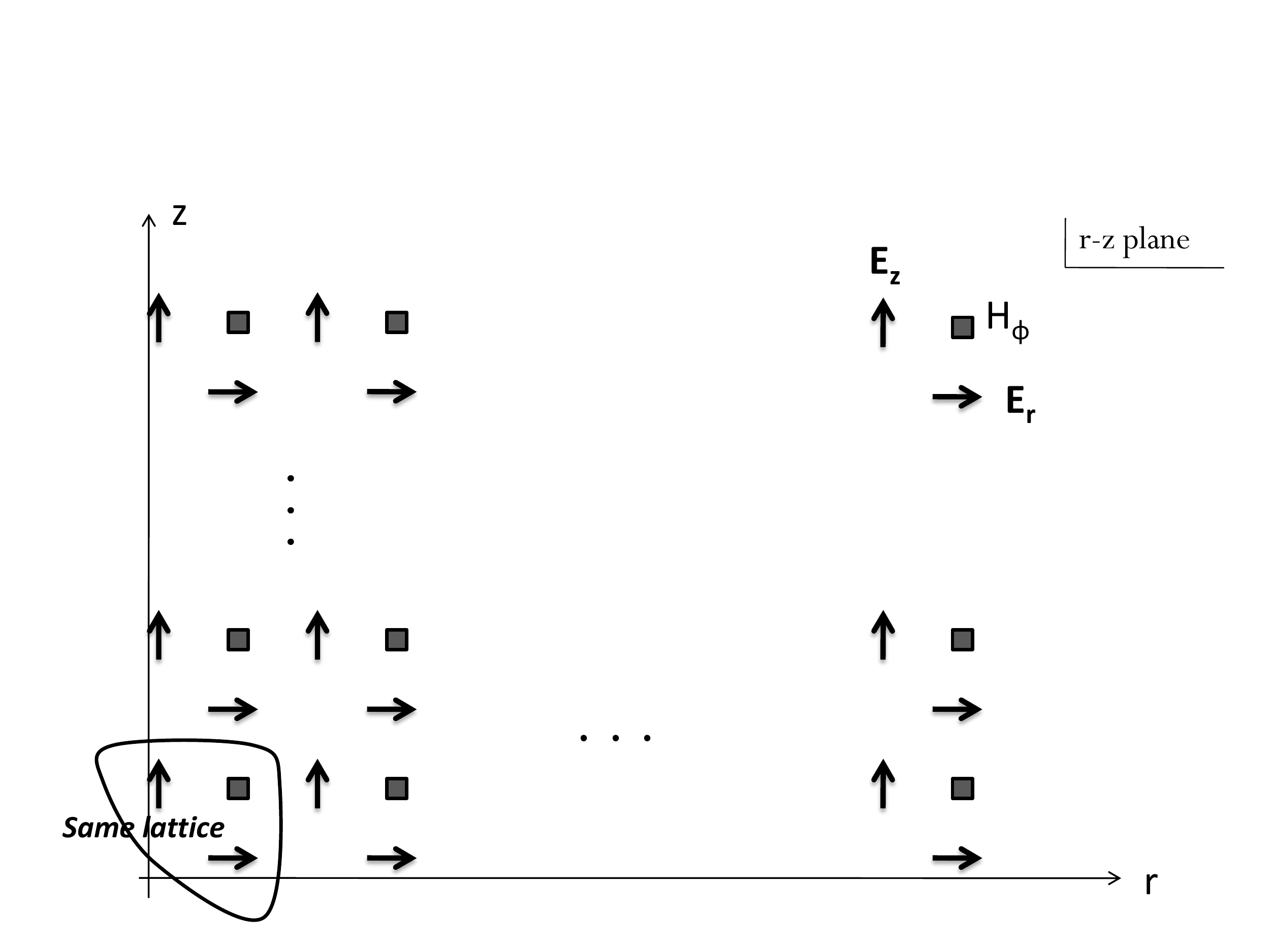}
	\caption{The adopted lattice configuration.}
	\label{fig:cylind_grid}
	\end{center}
\end{figure}

\subsection{Simulation Setup}
In order to not lose focus on the RF calculation, 
we simply assume a shower model for longitudinal development rather than to invoke the Monte Carlo packages.
For electromagnetic showers, 
the well known Nishimura-Kamata-Greisen (NKG) parameterization formula~\cite{NKG} that describes the number of shower particles reads
\small{
\begin{equation}
	N_e(X) = \frac{0.31}{ \sqrt{\ln{E_0/E_c}} } \exp \left[\left(1-\frac{3}{2}\ln{s}\right)\frac{X}{X_R}\right],
	\label{eq:nkg_formula}
\end{equation}
}
\normalsize
where $E_0$ is the energy of the primary particle and $E_c$ is the critical energy,
$X$ is the slant depth, $X_{max}$ the depth where the shower reaches its maximum number,
and $s$ is the (dimensionless) shower age defined as $s = 3X/(X+X_{max})$.
The NKG formula can be fit by a Gaussian distribution as:
\small{
\begin{equation}
	N_e(z) = N_{{\rm max}} \exp(-\frac{z^2}{2l^2}),
	\label{eq:showerModel}
\end{equation}
}
\normalsize
where $N_{{\rm max}}$ is the particle number at shower maximum
and $l$ is the longitudinal shower length.
For charge distribution in a snapshot,
we assume Gaussian distribution in both radial ($r$) and longitudinal ($z$) directions:
\small{
\begin{equation}
	n(z,r) = \frac{N_e}{(2\pi)^{1.5}\sigma_z\sigma_r^2} 
	\exp \left(-\frac{(z-X)^2}{2\sigma_z^2}\right) 
	\exp \left(-\frac{r^2}{2\sigma_r^2}\right),
\end{equation}
}
\normalsize
where $z$ and $r$ are the cylindrical coordinates in units of g/cm$^{2}$,
and $\sigma_z$ and $\sigma_r$ represent the standard deviation of the distribution in $z$ and $r$ directions, respectively.
The $\sigma_z$ is generally smaller than $\sigma_r$ and thus has no significant effect on the radiation pattern.
We adjust different relations among $R$, $l$, $\sigma_r$ and $\theta$ to produce radiation in different zones.
The value of $E_0$ only affects the overall normalization magnitude and is not our focus,
so we simply set $E_0 = 10^{19}$ eV throughout the study to obtain a reasonable scale of magnitude.
More realistic magnitudes can be obtained from Monte Carlo simulations of the shower.
Following the same spirit, we do not extract the number of net charged particles from the number of total charged particles (the normal ratio is about 0.2) in our model.
Our main focus is on the impact of these parameters on the shape of the Cherenkov radiation in both time-domain (the waveform) and frequency-domain (the spectrum).
The medium we use is ice, and the corresponding parameters are set accordingly.
That is, the critical energy $E_c$ is 73 MeV, the density of ice is 0.92 g/cm$^3$,
and the index of refraction is 1.78.

The lateral distribution $\sigma_r$ is set at 1m throughout the study for numerical purpose.
This is about ten times larger than the actual shower Moliere radius.
To avoid unmanageable computing time, our numerical choice of the Moliere radius is dictated by the resolution in FDTD method that we use. This compromise, however,
does not affect the physics we are discussing.
A more realistic situation should again rely on shower Monte Carlo simulations,
which would give the correct lateral size of showers,
and then use FDTD method with finer resolution to obtain the radiation pattern.
Of course, finer resolution requires more computing time and resources.
In principle, one can apply the scaling of each quantity in FDTD method. 
The scaling relationships between the grid size and the physical quantities in FDTD method are: if the grid size reduces by K times, then the time scale should also decrease by K times, while the frequency, the charge current, and the electric field should all increase K times.

Figure~(\ref{fig:setup}) is a cartoon that depicts our setup.
The shower travels in the +z direction emitting Cherenkov radiations.
We define a spherical coordinate whose origin lies in the intersection of the z axis and the shower maximum.
Desired detector positions are chosen by varying both the detection angle ($\theta$) and detection distance ($R$), 
and record the electric field values as time evolves.
After one single run, we were able to obtain all the simulation results we needed.
Here one can see the merit of using a time-domain calculation method.

\begin{figure}[htbp]
	\begin{center}
	\includegraphics[width=6.5cm]{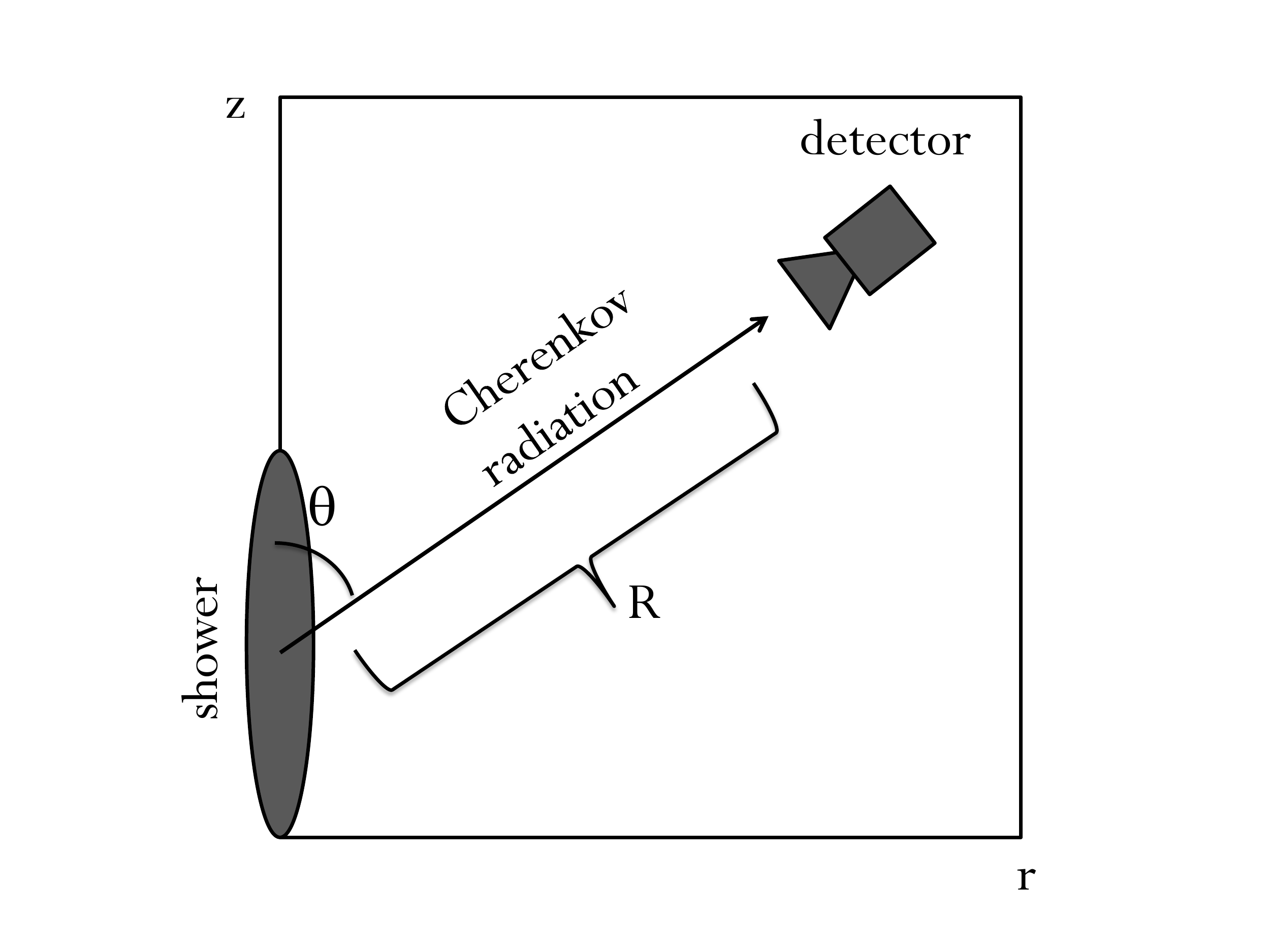}
	\caption{Schematic diagram for the simulation set up.}
	\label{fig:setup}
	\end{center}
\end{figure}

\subsection{Performance Improvement via the GPU Parallelization}
Graphical processing units (GPU) have became increasingly important in the field of high performance computation. 
Sparked by the needs of computer game markets, GPUs are currently advancing under booming developments.
They have become programmable via the Compute Unified Device Architecture (CUDA) provided by NVIDIA, 
and have been implemented in various scientific areas such as molecular dynamics, gravitational N-body simulations and lattice QCD
~\cite{MD, sapporo, cudanbody, gamer, lattQCD}.
CUDA is an extension of the C language, one of the most popular high-level languages in the world.
Programmers who are familiar with C language can utilize GPU computations by simply calling the functions from CUDA.
General parallelization strategies can be found in the CUDA Programming Guide available on their webpage.
Because of its multicore architecture, 
GPUs are ideal for implementing parallel algorithms.
The FDTD method is therefore an ideal candidate to benefit from GPU computing.
With the help of GPU parallelization,
we have obtained a great improvement on computing efficiency which otherwise would be so low that it makes the FDTD method formidable.

We run our code by a NVIDIA GTX285 graphic card provided by 
the Center for Quantum Science and Engineering of National Taiwan University (CQSE).
Table (\ref{tab:benchmark}) shows the improvement of GPU acceleration using a NVIDIA GTX285 graphic card,
compared to Intel Quad Core i7 920 at 2.66GHz (sequential code without CPU parallelization).
The computing time can be accelerated up to about 200 times,
depending on the total number of grids.
This is a great efficiency improvement, from more than a day to only ten minutes.

\begin{table}[h]
\begin{center}

\begin{tabular}{ l | c c c c } \hline \hline

$N_{\rm{grid}}$		&$t_{\rm{CPU}}$	&$t_{\rm{GPU}}$	&speed up		&GPU performance\\
			&(sec)  			&(sec)  			&(${t_{\rm{CPU}}}/{t_{\rm{GPU}}}$)  &(GFLOPS)\\
\hline
$256^2$		&5.15				&0.040				&128.75			&57.0 \\
$512^2$		&34.94				&0.186				&187.85			&98.1\\
$1024^2$	&294.41				&1.167				&252.28			&125.1\\
$2048^2$	&2122.80			&8.372				&253.56			&139.5\\
$4096^2$	&15751.17			&67.617				&232.95			&138.2\\
$8192^2$	&120617.48			&681.082			&177.11			&109.8\\
\hline \hline

\end{tabular}
\end{center}
\caption{Performance of CPU and GPU codes.}
\renewcommand{\arraystretch}{1.5}
\label{tab:benchmark}
\end{table}


\section{Results}

\subsection{At the Cherenkov Angle}
We first focus on radiation field right at the Cherenkov angle $\theta_c$.
The magnitude of the E-field, $E_\omega$, of course, decreases as the detection distance $R$ increases.
However, the scaling relation for such a decrease behaves quite differently between the near-field and the far-field regimes.
Figure (\ref{fig:near_match}) and Figure (\ref{fig:far_match}) show the spectrum of E-field at different distances $R$ with different scalings.
The longitudinal length $l$ is assumed to be 20 m,
which is roughly the length of a shower with $10^{18}$eV primary energy.

For $R$ = 50 m to 100 m, Figure (\ref{fig:near_match}) suggests that the magnitude of E-field goes like $1/\sqrt{R}$ in all frequency range.
On the other hand, for $R$ = 100 m to 200 m, the lower frequency part of the spectrum decreases faster than $1/\sqrt{R}$.
In fact, as Figure (\ref{fig:far_match}) implies,
the lower frequency part goes like $1/R$.
This phenomenon can be interpreted in a physical explanation.
It is a known fact that high frequency waves are less likely to diffract than low frequency ones.
As a result, the high frequency components of the Cherenkov radiation are more confined in the $\theta$-direction and their energies can only spread into the Cherenkov cone,
which makes them go like $1/R$ due to geometrical reason.
The fields are thus go like $1/\sqrt{R}$, which is a behavior of cylindrical wave.
For the low frequency components, 
diffraction allows another direction (the $\phi$-direction) for their energies to disperse,
which therefore makes them decrease faster.
As $R$ becomes large enough to satisfy the Fraunhofer limit~\cite{AZ_unified, AZ_thinned}, the radiation source can be viewed as a point,
which leads to a spherical wave behavior where energy goes like $1/R^2$.
In consequence, the scaling between $E_\omega$ and $R$ of all frequency range again has a simple relation: $E_\omega$ $\propto$ $1/R$.

\begin{figure}[H]
	\begin{center}
	\includegraphics[width=9cm, height=6cm]{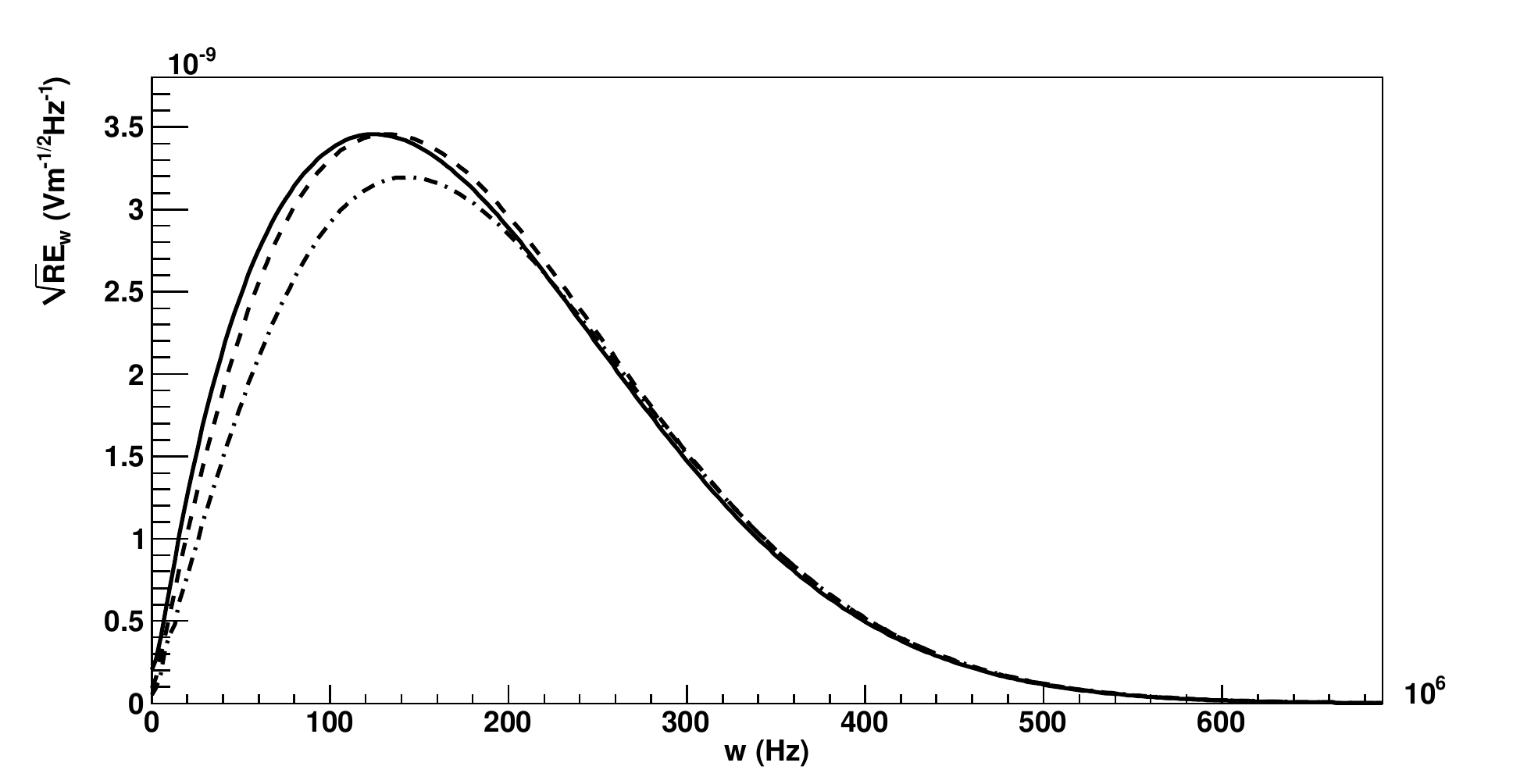}
	\caption{$\sqrt{R}|E_\omega|$ spectra at $R$ = 50 m, 100 m and 200 m (from top to bottom), for $l$ = 20 m and $\sigma_r$ = 1 m.
	These three spectra match well in thet high frequency regime,
	implying a cylindrical behavior ($E_\omega$ $\propto$ $1/\sqrt{R}$).}
	\label{fig:near_match}
	\end{center}
\end{figure}

\begin{figure}[H]
	\begin{center}
	\includegraphics[width=9cm, height=6cm ]{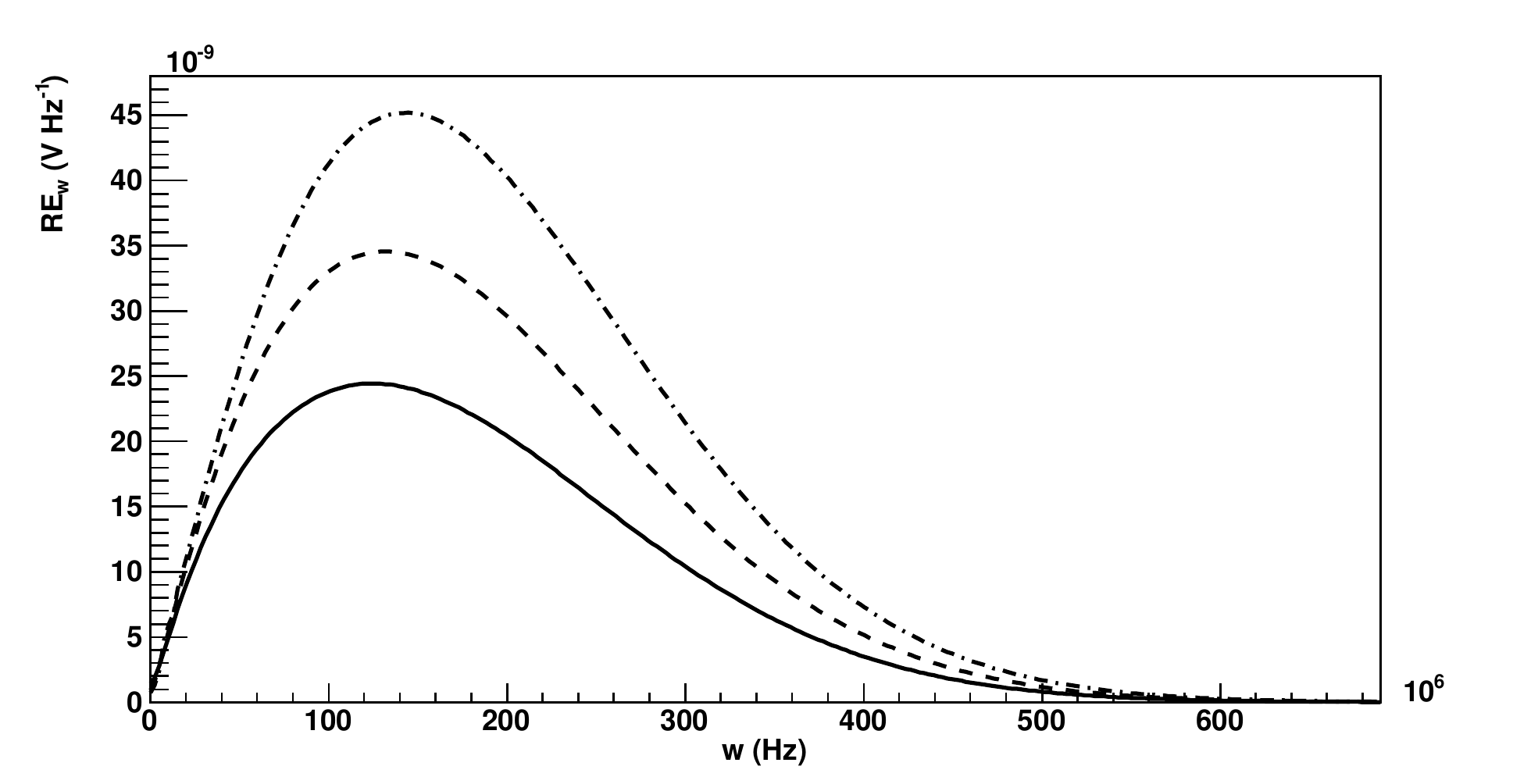}
	\caption{$R|E_\omega|$ spectra at $R$ = 200 m, 100 m and 50 m (from top to bottom), for $l$ = 20 m and $\sigma_r$ = 1 m.
	These two spectra match well in the low frequency regime,
	implying a spherical behavior ($E_\omega$ $\propto$ $1/R$).}
	\label{fig:far_match}
	\end{center}
\end{figure}

In Fresnel zone, the scaling behavior is a smooth transition from cylindrical to spherical.
Since components of different frequencies reach the Fraunhofer condition at different $R$,
the shape of the spectrum in Fresnel zone would change.
To be specific, higher frequency part meets Eq.(\ref{eq:farfield_condition}) at larger $R$,
and thus decreases slower in Fresnel zone.
The peak of the spectrum, as a result, 
would migrate to higher frequency as $R$ move further away.
This phenomenon can also be observed in time domain.
Figure(\ref{fig:waveform_200_to_700m}) shows the waveform from near-field to far-field.
There are two interesting features for the waveform.
First, the waveform in the near-field ($R$ = 200 m) has a small tail in its rear part, 
which gradually vanishes as $R$ increases.
Second, the waveform in the near-field is an asymmetric bipolar pulse in contrast to the symmetric bipolar pulse in the far-field.
These two features can be understood in a qualitative sense.

\begin{figure}[H]
	\begin{center}
	\includegraphics[width=9cm, height=6cm ]{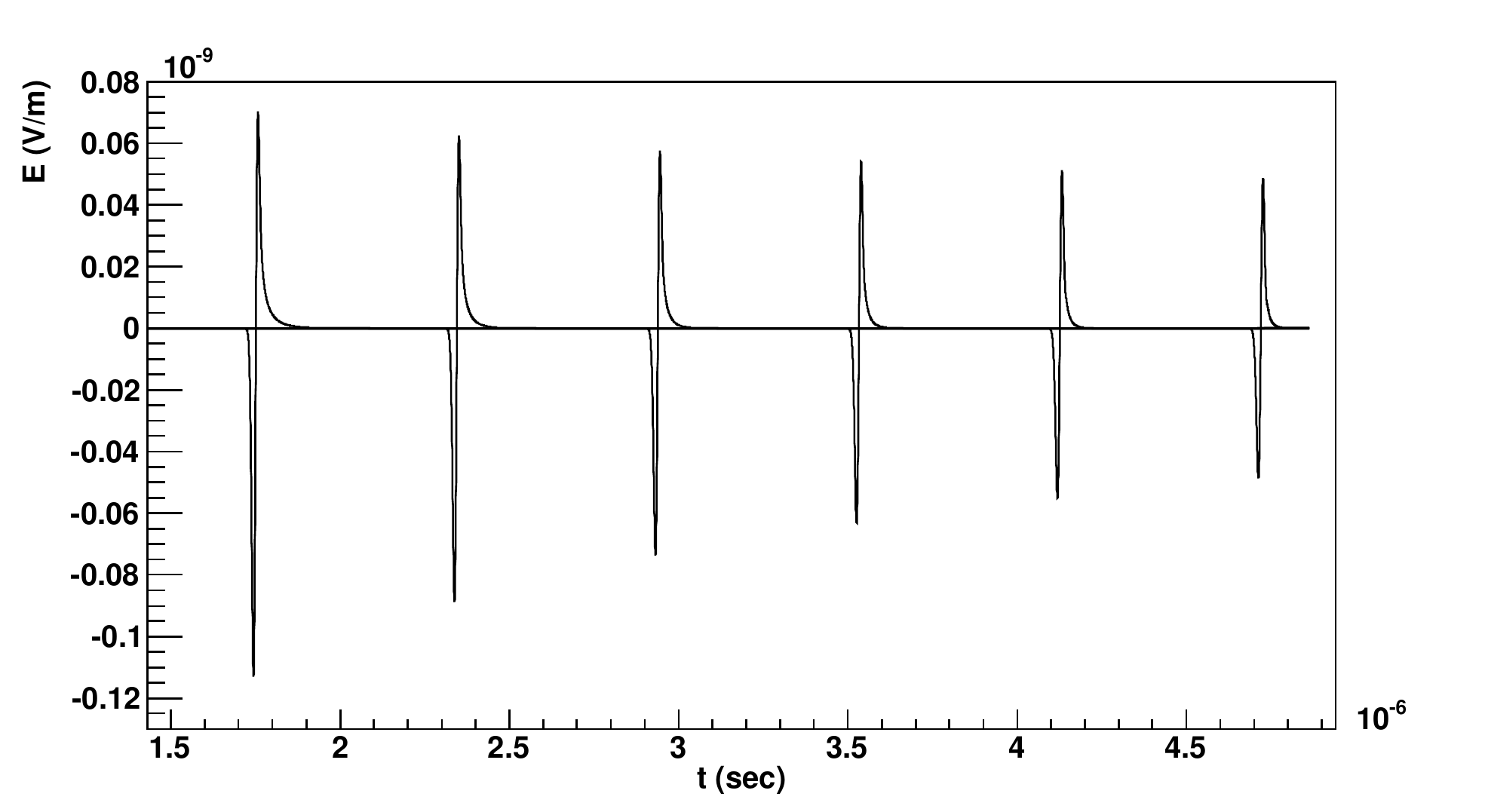}
	\caption{Time domain waveform at $R$ = 200, 300, 400, 500, 600, 700 m (from left to right), for $l$ = 50 m and $\sigma_r$ = 1 m. The tail part in the near-field is clear. Notice the transition of the asymmetric waveform in the near-field to the symmetric waveform in the far-field. }
	\label{fig:waveform_200_to_700m}
	\end{center}
\end{figure}

Figure (\ref{fig:first_light}) depicts a geometrical relation of Cherenkov radiation for a single charged particle emitted from different points.
Suppose an observer is at point $a$ and the shower travels along the z-axis,
wherein $z_0$ locates at the Cherenkov angle,
and $z_+$($z_-$) is any point lies right(left) to $z_0$.
Compare the arrival times of signals emitted from point $z_0$ and $z_+$.
When the charged particle passes $z_0$, it emits a signal.
Later on, as the shower passes $z_+$, it emits another signal.
The previous signal, in the mean time, has reached to point $b$.
Therefore with the simple triangular relation,
it should be clear that any signal emitted from the point right to $z_0$ would arrive the observer later than the signal emitted from $z_0$.
This is commonplace: the signal emitted first would end up arriving first.
However, the situation would be reversed on the other side.
When the shower passes $z_0$ and emits a signal,
the signal that emitted from $z_-$ has only traveled a distance that equals to the distance between $z_-$ and $c$, which means the signal has not even reached point $d$.
Therefore, signal emitted from the point left to $z_0$ would still arrive the observer later than the signal emitted from $z_0$.
It may seem counter-intuitive for a signal emitted later would in the end arrive earlier.
The crux of this unusual behavior is that the charged particle in ice travels faster than  electromagnetic wave does.
The signal emitted from $z_0$ is thus the first signal during the whole process.

With this in mind, we are now able to describe near-field radiation qualitatively.
When the first signal from $z_0$ arrives the observer,
it would generate a sudden rise of the scalar and vector potential.
The rise can be expected to be relatively strong for two reasons.
Firstly, $z_0$ is the shower maximum having a strongest source.
Secondly, since signals coming from the vicinity of $z_0$ would all be squeezed into a very short time interval and arrive at almost the same instant.
This "squeezing effect" can also be found in the case of the ordinary Lienard-Wiechert potential in the vacuum, where the most intensified direction points towards the traveling direction.
After the rise, the potential would experience a smoother decrease,
which results from the both sides of $z_0$ where signals would not be squeezed and the number of charged particle are fewer.
The corresponding E-field is just the derivative of the potential,
so the waveform should be a sharp peak coming first with a smooth tail following up in an opposite polarization.
A longer longitudinal length would generate a smoother tail.

\begin{figure}[H]
	\begin{center}
	\includegraphics[width=9cm, height=6cm ]{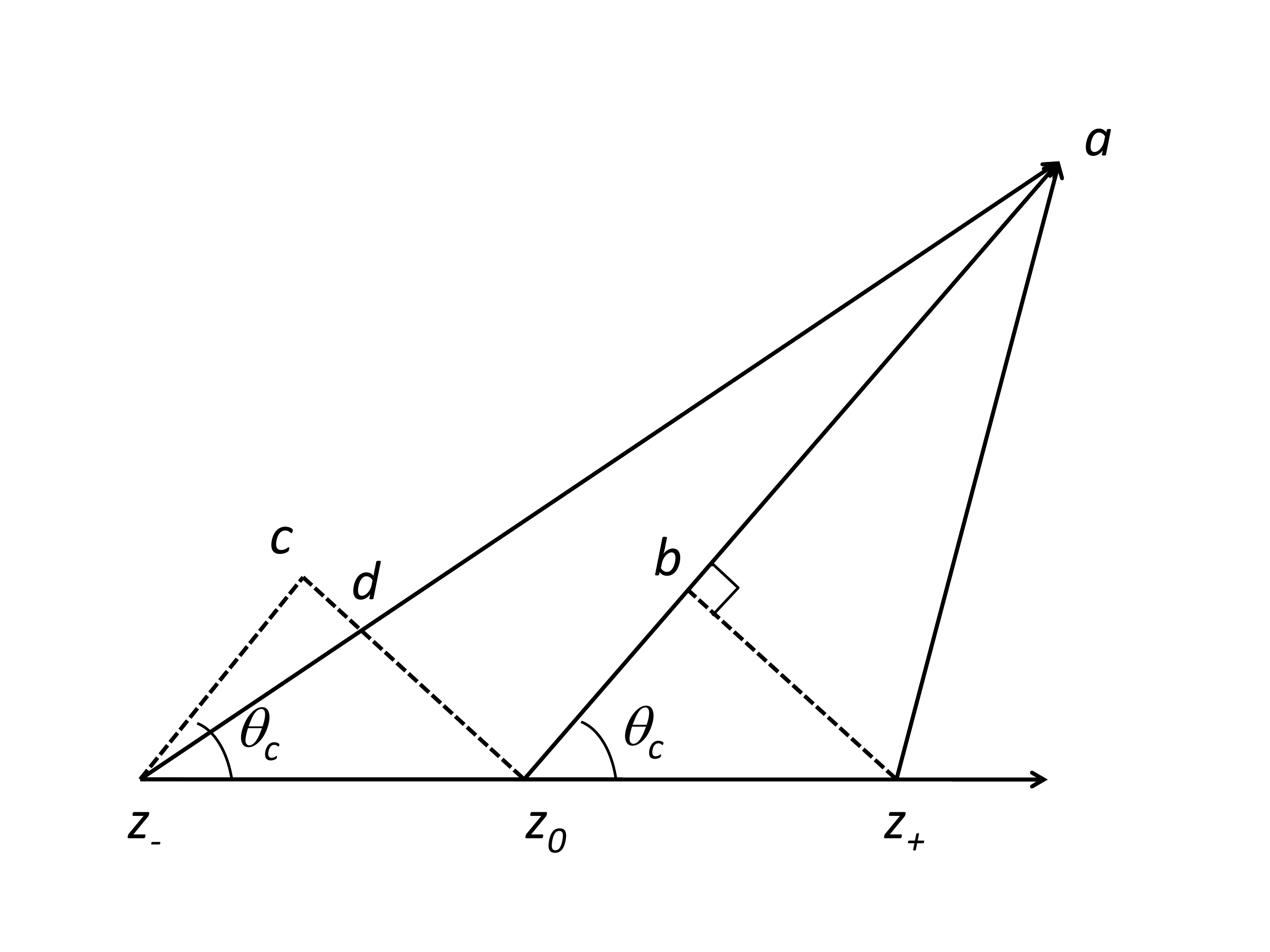}
	\caption{The geometrical relation of Cherenkov radiation for a single charged particle emitted from different points. The charged particle travels along z-axis and the observer is located at point $a$.}
	\label{fig:first_light}
	\end{center}
\end{figure}

In fact, the sharp peak is actually a delta function since the sudden rise in potential is represented by a step function.
A delta function waveform is unphysical and only appears in the one dimensional approximation which neglects the lateral distribution.
In real cases, the lateral distribution of the shower plays a part in interference and smooths the delta function.
Namely, the lateral distribution, based on its size, filters out high frequency components of the spectrum via destructive interference.
As the detection distance moves further away,
signals from the whole range of longitudinal development arrive at the same time.
In consequence, the potential decreases drastically after the sudden rise.
For $R$ large enough (Fraunhofer limit),
the potential vanishes as fast as its rise,
and thus resembles a delta function.
Therefore, its derivative leads to a bipolar waveform symmetric in time domain.
This is the reason the asymmetric waveform in near-field gradually transform into a symmetric one as $R$ increases.
The peak shift of the spectrum can be understood as the longitudinal development gradually lose its impact on the waveform as $R$ increases.
In this case, 
the characteristic frequency of the waveform is determined uniquely by the interference of the lateral distribution.


\subsection{Angular Distribution}

We now start to investigate the radiation for a fixed $R$ at different angles.
Figure (\ref{fig:spec_fresnel}) shows the near-field spectrum for $R$ = 300 m, $l$ = 20 m and $\sigma_r$ = 1 m.
We compared the result of FDTD method with the one of saddle-point approach and found they are in good agreement.
The peak frequency is highest at the Cherenkov angle and decreases as $\theta$ moves away from $\theta_c$, which is the main feature of diffraction.

\begin{figure}[H]
	\begin{center}
	\includegraphics[width=9cm, height=6cm ]{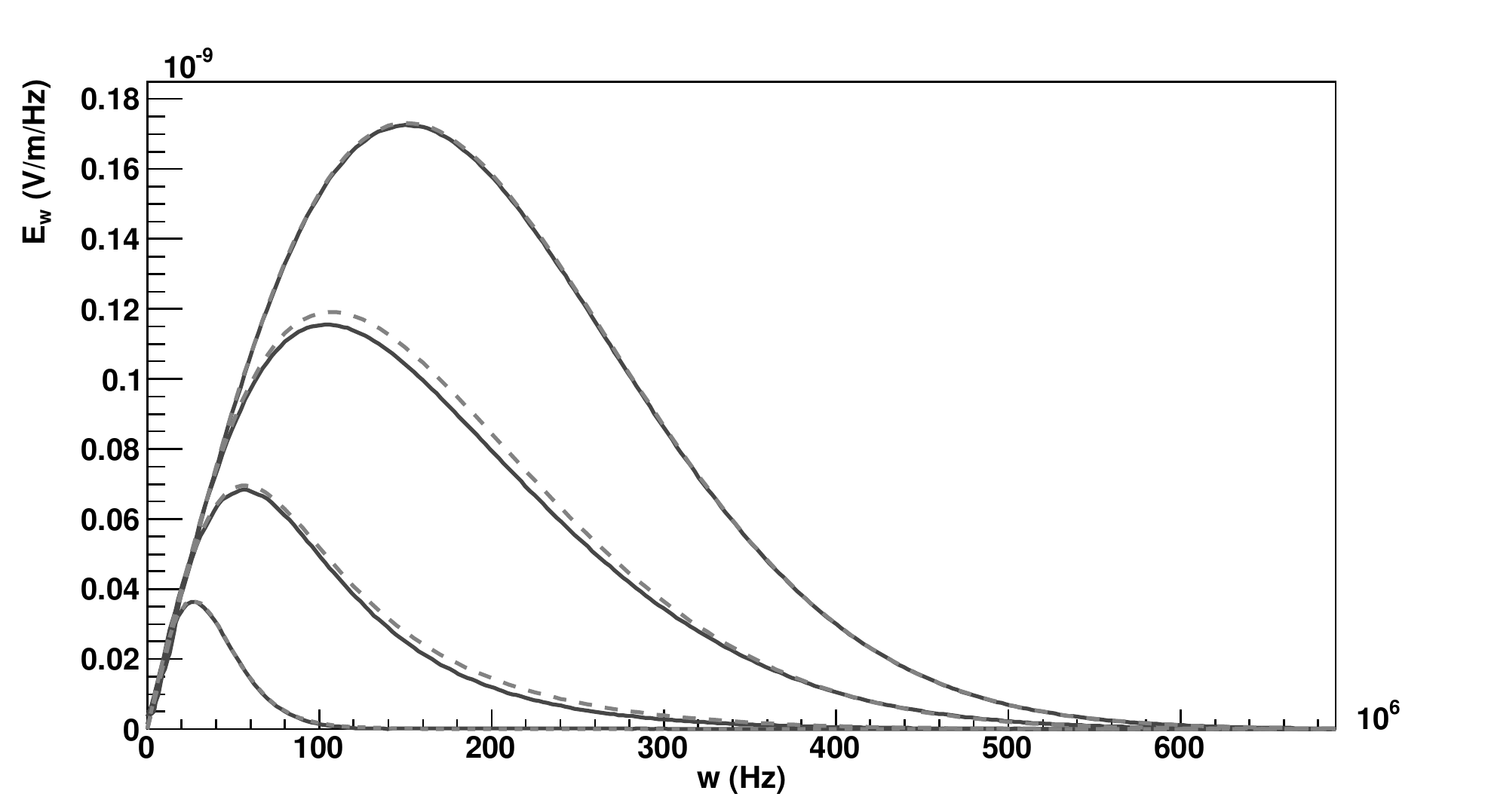}
	\caption{Fresnel zone $E_\omega$ spectra at $\theta$ = $\theta_c$, $\theta_c$ + 5$^{\circ}$, $\theta_c$ + 10$^{\circ}$ and $\theta_c$ + 20$^{\circ}$ (from top to bottom), for $R$ = 300m, $l$ = 20 m and $\sigma_r$ = 1 m. The solid curves are our results and the dashed curves are obtained by the saddle-point approach. Results from both methods are in good agreement. The peak frequency is highest at the Cherenkov angle.}
	\label{fig:spec_fresnel}
	\end{center}
\end{figure}

Figure (\ref{fig:spec_near}) shows the near-field spectrum  at different angles for $R$ = 300 m, $l$ = 100 m and $\sigma_r$ = 1 m.
The disagreement between the two methods is obvious since the saddle-point approach is not suitable in the near-field regime.
An interesting feature is that at different angles the peak frequency does not change much.
The reason is simple:
in the near-field the detection distance is too short for the diffraction to take effect, 
and the radiation only exists within the area sweeping by the longitudinal length along $\theta_c$.
The peak magnitude at different angles reflects the longitudinal development.
As the radiation travels into the Fresnel zone or even Fraunhofer zone,
the diffraction effect gradually appears and part of radiation (especially lower frequency components) leaks out of the Cherenkov angle,
as illustrated in Figure(\ref{fig:near-field sweeping zone}).


One thing worth noticing is the peak frequency at $\theta_c$ in Figure (\ref{fig:spec_near}) is lower than that in Figure (\ref{fig:spec_fresnel}).
It is an example of how the longitudinal length affecting the characteristic frequency at the Cherenkov angle.
We can expect a long tail in the waveform in such case.

\begin{figure}[H]
	\begin{center}
	\includegraphics[width=9cm, height=6cm ]{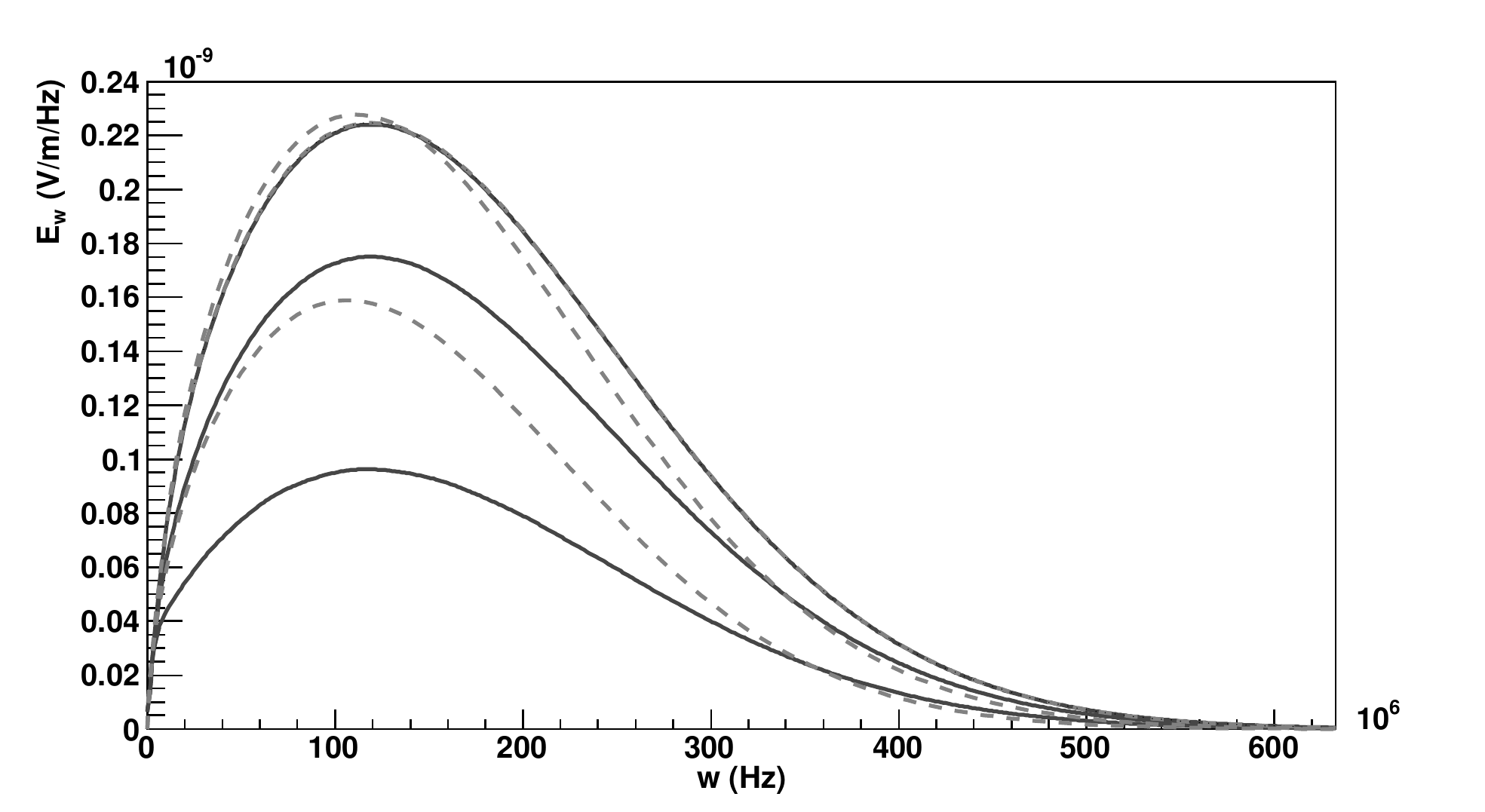}
	\caption{Near-field $E_\omega$ spectra at $\theta$ = $\theta_c$, $\theta_c$ + 10$^{\circ}$ and $\theta_c$ + 20$^{\circ}$ (from top to bottom), for $R$ = 300m, $l$ = 100 m and $\sigma_r$ = 1 m. The solid curves are our results and the dashed curves are obtained by the saddle-point approach, which is not suitable in the near-field. The peak frequency is almost a constant at different angles. }
	\label{fig:spec_near}
	\end{center}
\end{figure}

We now turn to the angular distribution for components of different frequencies.
Figure (\ref{fig:ang_fresnel}) shows the angular distribution in Fresnel zone for $\omega$ = 50, 100 and 200 MHz respectively for  $R$ = 300 m, $l$ = 20 m and $\sigma_r $ = 1 m.
The diffraction pattern is clearly shown:
higher frequency component has a narrower angular distribution.
All three curves are symmetric in both sides of the Cherenkov angle.

In near-field, the situation would be quite different.
Figure (\ref{fig:ang_near}) shows the angular distribution in near-field for $\omega$ = 20, 50 and 100 MHz respectively for  $R$ = 300 m, $l$ = 20 m and $\sigma_r $ = 1 m.
Radiation is stronger at $\theta < \theta_c$ than the other side,
since the far-field approximation replaces $|\vec{x}-\vec{x}\,'|$ with a constant $R$ in the denominator and therefore neglects the fact that the region $\theta < \theta_c$ lies closer to the shower axis and would surely have a larger signal.
For a large enough $R$,
since the whole longitudinal development can be viewed as located at the same point,
this difference becomes negligible,
resulting in a symmetric angular distribution.

\begin{figure}[H]
	\begin{center}
	\includegraphics[width=9cm, height=6cm ]{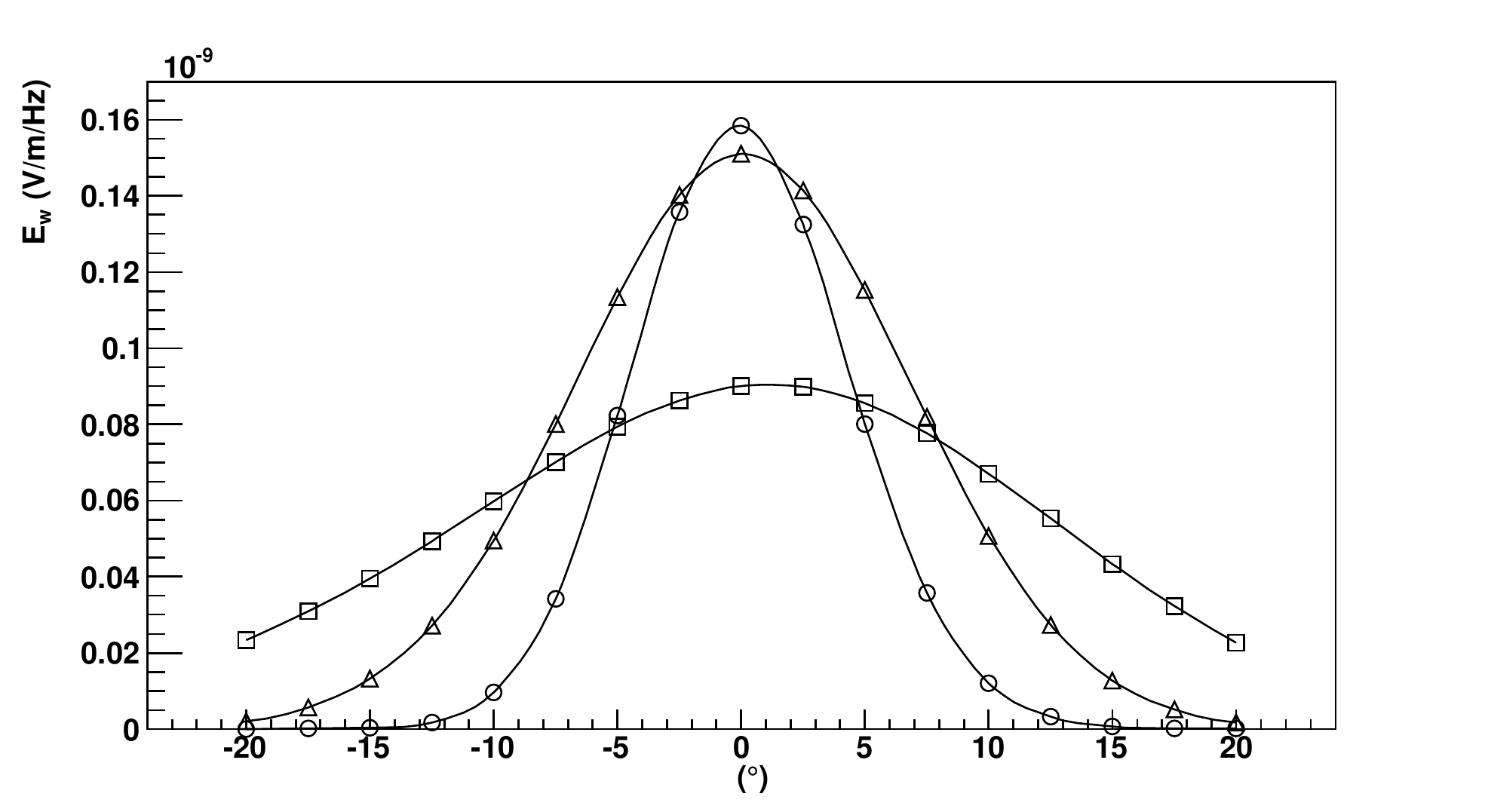}
	\caption{Fresnel zone angular distribution for $\omega$ = 200, 100, 50 MHz (from top to bottom) with $R$ = 300 m, $l$ = 20 m and $\sigma_r $ = 1 m. We redefine $\theta$ = 0$^{\circ}$ as the Cherenkov angle for clarity. The angular distribution is symmetric at both sides. The diffraction pattern is presented clearly.}
	\label{fig:ang_fresnel}
	\end{center}
\end{figure}

\begin{figure}[H]
	\begin{center}
	\includegraphics[width=9cm, height=6cm ]{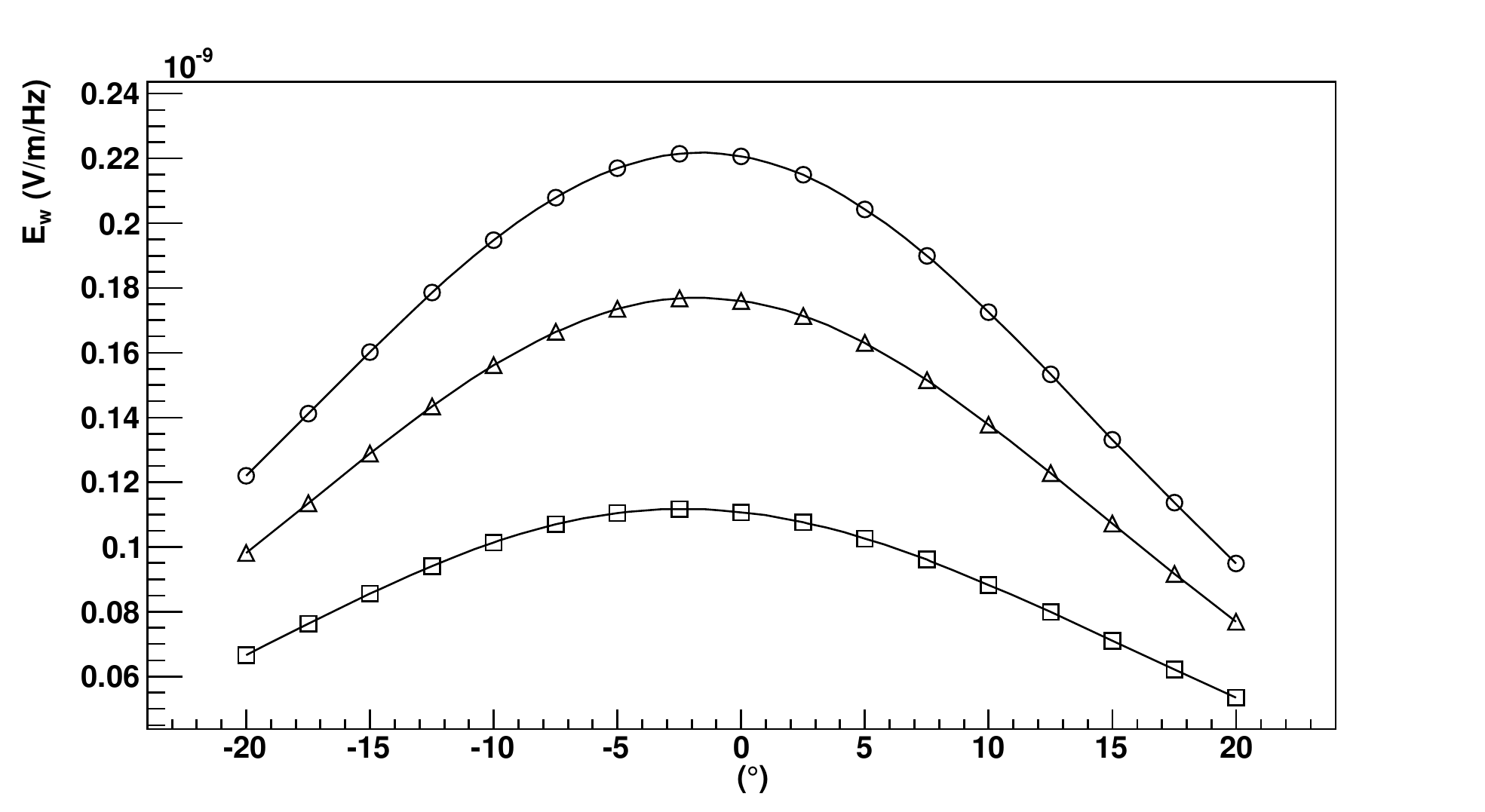}
	\caption{Near-field angular distribution for $\omega$ = 100, 50, 20 MHz (from top to bottom) with $R$ = 300 m, $l$ = 100 m and $\sigma_r $ = 1 m. We redefine $\theta$ = 0$^{\circ}$ as the Cherenkov angle for clarity. It shows an asymmetric angular distribution at both sides.}
	\label{fig:ang_near}
	\end{center}
\end{figure}

Time domain waveform also gives some interesting properties.
Figure (\ref{fig:waveform_near}) shows the waveform in near-field at different angles.
As mentioned above,
the near-field waveform is asymmetric in time.
The characteristic frequency can be implied by the duration of the pulse.
The duration of the pulse at different angles changes quite little,
which we have already seen in Figure(\ref{fig:spec_near}).
One intriguing phenomenon is the arrival time difference of the waveform.
This subtlety is due to the fact that the near-field wavefront is a straight line on the r-z plane instead of an arc in the Fraunhofer limit.
Consequently, radiation arrives last to the observer at $\theta$ as it can be seen in Figure (\ref{fig:waveform_near}).

In Fresnel zone,
as shown in Figure (\ref{fig:waveform_fresnel}),
the arrival time difference vanishes if we take the zero-crossing instant as the arrival time.
The waveform becomes symmetric at $\theta_c$ while it remains asymmetric at other angles.
As mentioned earlier,
symmetric waveform results from the "squeezing effect",
which leads to a rapid decay of potential after arrival of the first signal.
For $\theta$ away from $\theta_c$,
the most squeezed region ($z_0$) correspond to an area where either the shower has not yet started to develop ($\theta > \theta_c$) or the shower development is over ($\theta < \theta_c$).
The signals emitted from the whole shower in this case would arrive in order,
either ordinary ($\theta > \theta_c$) or reversed ($\theta < \theta_c$),
which leads to a smooth decay of the potential and thus an asymmetric waveform.

\begin{figure}[H]
	\begin{center}
	\includegraphics[width=9cm, height=6cm ]{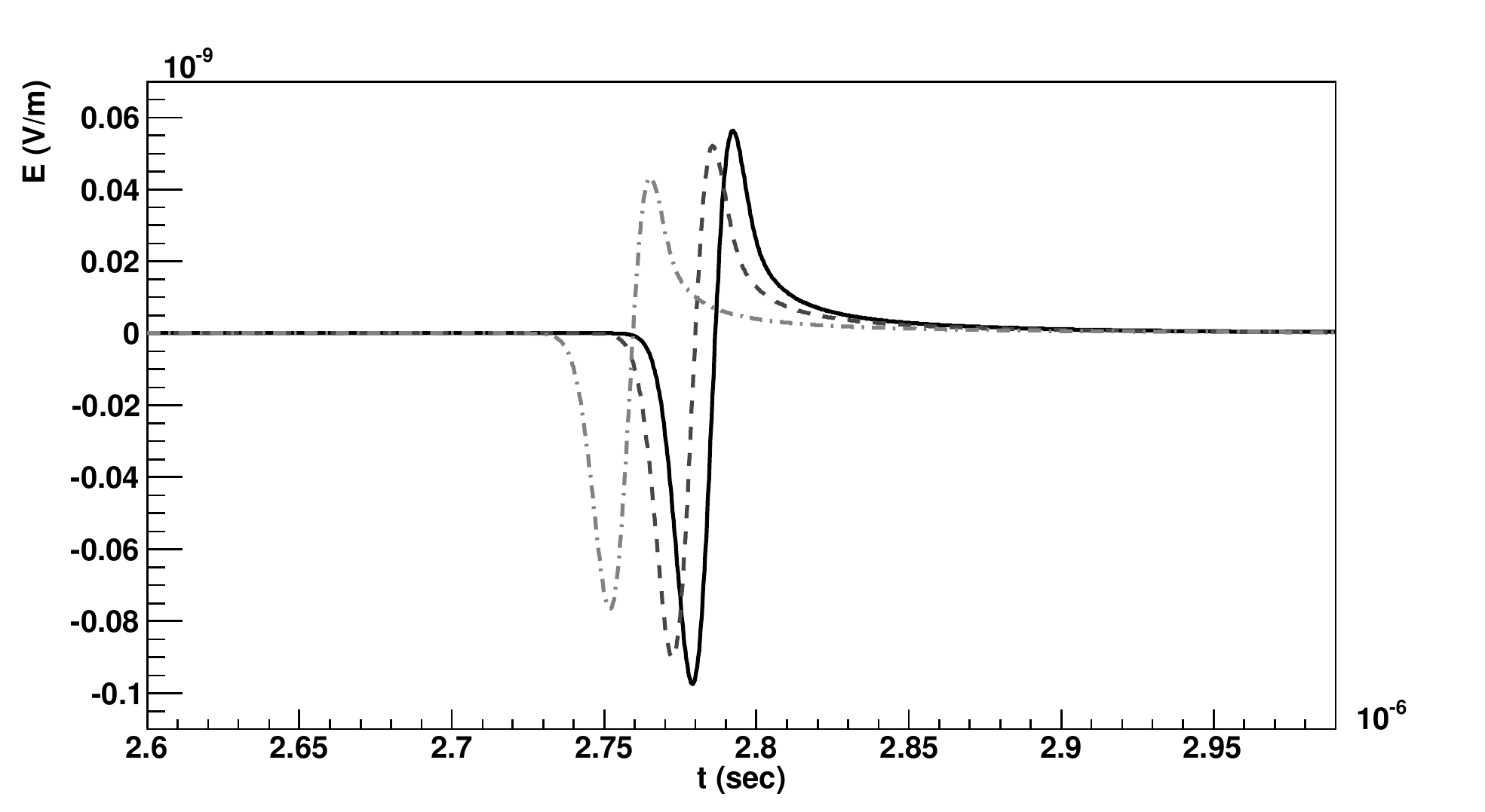}
	\caption{Waveform in near-field at different angles, for $R$ = 300m, $l$ = 100 m and $\sigma_r$ = 1 m. The solid curve is at $\theta$ = $\theta_c$, dashed curve is at $\theta$ = $\theta_c$ + 5$^{\circ}$ and dashed-dotted curve is at $\theta$ = $\theta_c$ + 10$^{\circ}$. The asymmetric waveforms are the striking feature of near-field radiation. Notice the arrival-time difference of the three waveforms. }
	\label{fig:waveform_near}
	\end{center}
\end{figure}

\begin{figure}[H]
	\begin{center}
	\includegraphics[width=9cm, height=6cm ]{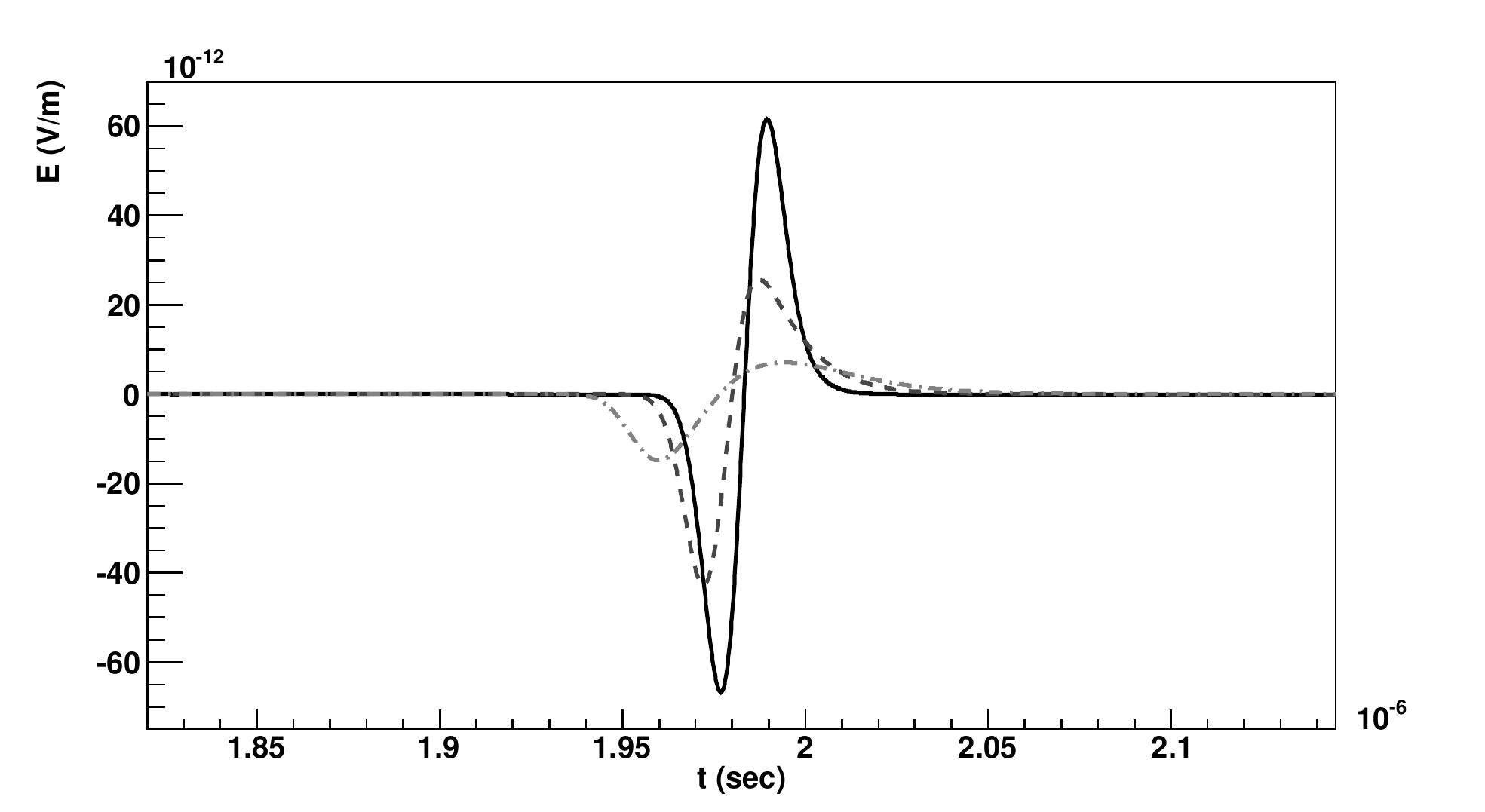}
	\caption{Waveform in Fresnel zone at different angles, for $R$ = 300m, $l$ = 20 m and $\sigma_r$ = 1 m. The solid curve is at $\theta$ = $\theta_c$, dashed curve is at $\theta$ = $\theta_c$ + 5$^{\circ}$ and dashed-dotted curve is at $\theta$ = $\theta_c$ + 10$^{\circ}$. The waveform becomes symmetric at $\theta_c$ and remains asymmetric at $\theta_c$ + 5$^{\circ}$ and $\theta_c$ + 10$^{\circ}$. }
	\label{fig:waveform_fresnel}
	\end{center}
\end{figure}


\section{Summary and Conclusions}

We have studied the near-field radiation in both time domain and frequency domain.
We shown that even for a shower with symmetric longitudinal development (e.g. Gaussian distribution),
the resulting near-field waveform would be asymmetric in time.
The longitudinal development is as important as the lateral distribution even at the Cherenkov angle.
Future work on the parameterization of near-field radiation should take this into account.

As the radiation propagates,
the waveform would gradually become symmetric.
Moreover,
this transition occurs at different $R$ for different $\theta$.
To be specific, it occurs at shortest distance for the observer located at the Cherenkov angle.

For a ground array neutrino detector,
with the size of LPM-elongated showers becomes comparable with the typical detection distance,
the near-field effect is an indispensable factor.
The correct relation of distance dependence in near-field prevents underestimation of the signal strength in a Monte Carlo simulation.
Furthermore, the correct angular spread in near-field is necessary in order not to underestimate the detection solid angle of a neutrino detector.
The Fraunhofer approximation leads to an angular spread that is quite narrow for LPM-elongated showers.
It is incorrect since a shower of hundred meters long would at least generate radiation which also spans hundred meters long in near-field.
The overall detector sensitivity should be better than adopting traditional radiation formula in Fraunhofer limit in the Monte Carlo simulation.
We plan to find the parameterization formula suitable in all cases in the future work.

On the other hand,
due to the complicated features of near-field radiation,
new reconstruction methods are required.
For example, the normal way to reconstruct the direction of incoming Cherenkov pulses is by the arrival time differences between antennas.
This method treats the wavefront as a spherical one and is based on the far-field assumption which fails in near-field as we have shown.

Charged-current interaction of electron neutrino is the main source of ultra-high energy electromagnetic showers.
In a typical detection distance,
the near-field condition should be easily satisfied.
Identification of such radiation therefore implies electron neutrino events.
This opens an opportunity for neutrino flavor identification,
since muon and tau neutrinos only induce hadronic showers whose sizes are too compact to produce near-field radiation in a typical detection distance.


\section{Acknowledgements}
We thank Ting-Wai Chiu for the help in GPU-calculations.
This research is supported by the Taiwan National Science Council(NSC) under Project No. NSC98-2811-M-002-501, No. NSC98-2119-M-002-001, 
the Center for Quantum Science and Engineering of National Taiwan University(NTU-CQSE) under Nos. 98R0066-65, 98R0066-69, 
and the US Department of Energy under Contract No. DE-AC03-76SF00515.
We would also like to thank the Leung Center for Cosmology and Particle Astrophysics and the Taiwan National Center for Theoretical Sciences for their support.

\end{document}